\documentclass[12pt,aps,prb,preprint]{revtex4}
\usepackage{amsmath}
\usepackage{graphicx}
\usepackage{subfig}

\begin{document}
\title{Coulomb Field Scattering in Born-Infeld Electrodynamics}
\author{Daniel Tennant}
\email{dtennant@austincc.edu} \affiliation{Department of Physics, Austin Community College, 78758}

\date{\today}

\begin{abstract}
In the context of Born-Infeld electrodynamics, the electromagnetic fields interact with each other via their non-linear couplings.  A calculation will be performed where an incoming electromagnetic plane wave scatters off a Coulomb Field in the geometrical optics approximation.  In addition to finding the first order angle of deflection, exact solutions for the trajectory will also be found.  The possibility of electromagnetic bound states will be discussed.
\end{abstract}

\maketitle


\section{Introduction}

Non-linear electrodynamics has been a subject of research for many
years.  That quantum electrodynamics predicts that the
electromagnetic field behaves non-linearly through the presence of
virtual charged particles\cite{euler} was first discussed by
Heisenberg and Euler.  This was rapidly followed by the proposal of
Born and Infeld (BI)\cite{born} of a new classical non-linear theory of
electromagnetism which possesses a maximum field strength, $\beta$.

The theory of BI has found a resurgence of attention due to its reproduction through string theory given certain boundary conditions\cite{frad}.  The BI theory also contains many symmetries common to the Maxwell theory despite its non-linearity.  The pioneering work of Plebanski\cite{pleb} and Adler\cite{adler} amongst others\cite{boil}\cite{bial} showed the BI theory to be unique among non-linear theories of electrodynamics in that it is the only possible theory that is absent of birefringence; both polarization modes propagate at the same velocity.    

More recently, it was observed that the BI theory has another common characteristic with (source-free) Maxwell.  BI is the only possible non-linear version of electrodynamics that is invariant under electromagnetic duality transformations \cite{gibb}.  Note that there is no physical demand that electrodynamics must be a linear field theory.  Therefore, the non-linear extensions of electrodynamics which respect the same symmetries must be taken into consideration. 

Unfortunately, the fact that it is invisible to optical rotation experiments such as PVLAS \cite{pvlas} makes it difficult to find experimental signatures of BI.  Recent experimental proposals include using waveguides \cite{ferraro} and ring lasers\cite{denisov}.  In this article, I will propose using a strong electric field to ``trap" an incoming electromagnetic wave.

Our incoming ray will behave as if it is in a novel space-time geometry.  In this article, the mechanism for establishing the non-trivial metric is considering electrodynamics to be a non-linear theory.  An incoming electromagnetic wave will experience a spatial and frequency dependant index of refraction which will cause a deflection in trajectory.  This index of refraction will have the equivalent effect as the incoming ray being in a warped space-time geometry.

The geometrization of space-time is due to the universality of gravity; it couples to all known forms of matter/energy.  Recently different analogue models of general relativity have appeared in hopes of detecting Hawking Radiation\cite{visser}.  In this article, another example of an effective geometrization of space-time will be presented.  Recall that what I mean by effective geometrization is that it is not a universal effect.  In our case, this effective geometry will be perceived only by electromagnetic plane waves.  Other forms of matter and energy will behave as in their usual space-time geometry.  

The special case we will consider is a Coulomb Field in the context of Born-Infeld electrodynamics.  This case actually has many similarities with the Schwarzschild solution of General Relativity.  We will find that they share a potential energy with some common characteristics.  The most interesting is the existence of a point of unstable equilibrium about which rays can be momentarily trapped.

\section{Wave Propagation}

Consider an incoming electromagnetic wave propagating through a background electromagnetic field.  One of the main difficulties with non-linear theories is that, in general, it is impossible to disentangle fields into separate parts (this part of the total field is due to this source, that part of the field is due to that source, etc.).  The ability to superimpose solutions to differential equations is a property of linear differential equations only.  

It is necessary to impose some conditions on our example to observe the behaviour of the incoming wave separate from the background field.  First, the incoming disturbance, $f_{ab}$, is ``weak" compared to the background field, $F_{ab}$.  This guarantees that the background field will not be affected by the incoming field.  A similar situation arises in General Relativity where one usually assumes that the mass moving in a background gravitational field does not affect that field.

Second, we can only consider incoming waves whose wavelengths that are sufficiently small.  To be precise, the background field cannot change appreciably over a wavelength.  This allows us to consider the background field to be approximately constant when compared to the incoming wave so that when we find the rate of change of the total field, it effectively isolates the wave portion, $\partial F_{T}\sim\partial f$.

Once these approximations are made, it is possible to write an expression for the disturbance propagating through a background field.  Combining the dynamical ``Maxwell" equations,
\begin{equation}
\partial_{b}\frac{\partial\mathcal{L}}{\partial F_{ab}}=0,
\end{equation}  
with the kinematic equations,
\begin{equation}
\partial_{a}f_{bc}+\partial_{c}f_{ab}+\partial_{b}f_{ca}=0,
\end{equation}
one arrives at
\begin{eqnarray}\label{wave}
&\partial^{2}f_{ab}+\Lambda^{rs}_{ab}f_{rs}=0.
\notag \\
&\Lambda_{ab}^{rs}=\frac{1}{2\mathcal{L}_{F}}[\mathcal{L}_{FF}F_{[a,}\,^{m}F^{rs}
+\mathcal{L}_{FG}(F_{[a,}\,^{m}\tilde{F}^{rs}+\tilde{F}_{[a,}\,^{m}F^{rs})+
\mathcal{L}_{GG}\tilde{F}_{[a,}\,^{m}\tilde{F}^{rs}]\partial_{b]}\partial_{m}
\end{eqnarray}
Brackets in the subscript denote the antisymmetric combination.  Terms such as $\mathcal{L}_{FG}$ are to be understood as $\frac{\partial^{2}\mathcal{L}}{\partial\mathcal{F}\partial\mathcal{G}}$.  Note that this expression holds for any Lagrangian density and any slowly-varying background field.

Following the usual prescription to arrive at the geometrical optics
regime\cite{landau}, the incoming wave is of the form $f\sim e^{iS}$,
where higher derivatives of the phase are negligible, $\mid\partial
S\mid^{2}>>\\ \mid\partial^{2}S\mid$.  This allows us to identify the
first partial derivatives of the field as $\partial_{a}S\rightarrow
k_{a}$.  

\section{Background Coulomb Field}

By the method of effective geometries laid out by Novello \cite{novello} and
others, I will describe a ray passing through a background Coulomb
field, $\frac{\mu}{r^2}\hat{r}$.  Due to the effect on the ray by the background field, the ray will no longer follow null geodesics in Minkowski space, but in another, effective geometry;
\begin{equation}\label{coulmet}
g^{ab}k_a k_b=\frac{1}{1-(\frac{\alpha}{r^2})^2}(-k^2_t+k^2_r)+r^2 k^2_\phi=0
\qquad \alpha=\frac{\mu}{\beta}. 
\end{equation}

Our coordinate system is arranged such that the incoming ray falls in the x-y plane.  Due to the symmetries inherent in this metric, one can expect to find a conserved energy, $\omega$, and angular momentum, $\delta$.  The effective metric can be expressed as a sum of the ray's kinetic and potential energies.
\begin{equation}
\dot{z}^2+U(z)=0 \qquad 
U(z)=\frac{\omega}{z^2}(1-(\frac{\gamma}{z^2})^2)(1-z^2+(\frac{\gamma}{z})^2)
\qquad \gamma=\frac{\alpha}{b^2}=\frac{\vec{E}(b)}{\vec{\beta}}
\end{equation}
The coordinate $z$ is the radial length in terms of the scattering parameter, $b=\frac{\delta}{\omega}$.  There is one local maximum for this potential.
\begin{eqnarray}
& z_o=-\frac{2}{3}[2\gamma^2+(a^2+b^2)^{1/6}\rm{cos}(\frac{\phi+2\pi}{3})] 
\qquad \textit{a}=(2\gamma)^6 \qquad \textit{b}=(3\gamma)^{3}[1+\frac{3}{4}(1+\frac{32}{9}\gamma^2)^2]^{1/2} \notag \\
& \rm{tan}\phi=\frac{\textit{b}}{\textit{a}}
\end{eqnarray}
This point ranges from approaching zero in the limit of no non-linear coupling to $z_o\simeq 1.23$ in the opposite limit, $\gamma\rightarrow 1$.

By analysing the potential, we can ascertain what types of orbits are possible.  For the most physically realistic case, $\gamma<<1$, the potential is slightly repulsive.  However, at approximately $\gamma\simeq 0.87$, the incoming ray passes close enough to cross the potential maximum such that attractive and repulsive forces cancel and the ray continues on its original trajectory.  For gamma values greater than this, the ray is bent inward.  We will see below that for $\gamma=1$ the ray is actually caught in a bound, unstable orbit.  

\begin{figure}
  \centering
  \subfloat[$\gamma=0.2$]{\includegraphics[width=0.3\textwidth]{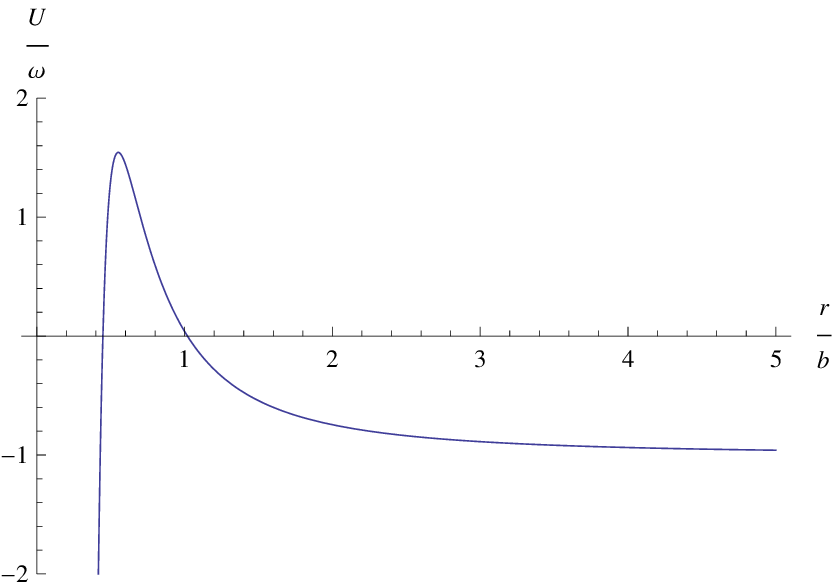}}                
  \subfloat[$\gamma=0.87$]{\includegraphics[width=0.3\textwidth]{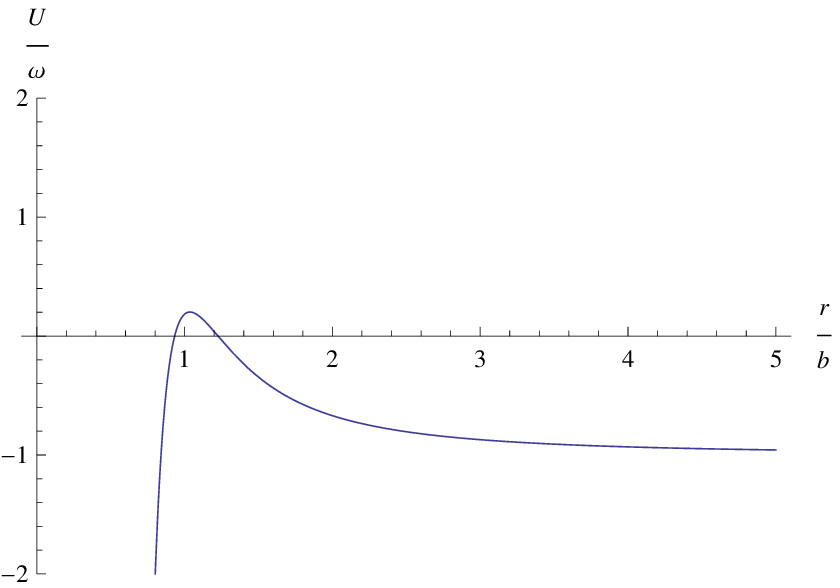}}
  \subfloat[$\gamma=0.95$]{\includegraphics[width=0.3\textwidth]{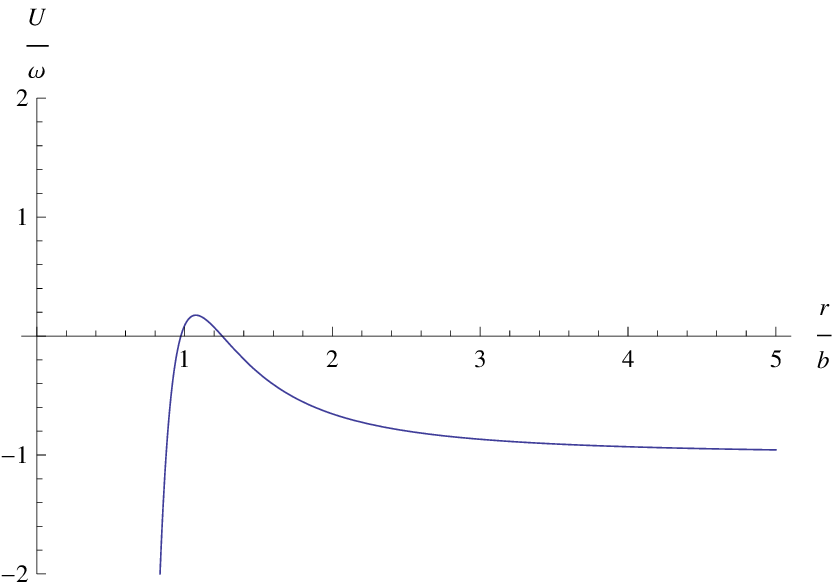}}
  \caption{Potential Energy}
\end{figure}

Exact solutions for the trajectory in this effective geometry are attainable in terms of elliptical functions.  It is necessary to make some coordinate transformations; $x=\frac{r^{2}}{\alpha}$ and $\phi=\frac{1}{2}\varphi$.  Equation (\ref{coulmet}) becomes
\begin{equation}\label{new}
(\frac{dx}{d\varphi})^2=\gamma(x-1)(x+1)(x-1/\gamma)\equiv f(x).
\end{equation}

Physically possible trajectories correspond to positive values of $f(x)$.  Again, the most physically relevant case is that of the incoming ray with $\gamma<<1$.  In this case, $x$ approaches $1/\gamma$ from infinity and then returns out to infinity.  This corresponds to the ray approaching from infinity, reaching $b$, the impact parameter, and continuing on.  

\begin{figure}
  \centering
  \subfloat[$\gamma=0.2$]{\includegraphics[width=0.3\textwidth]{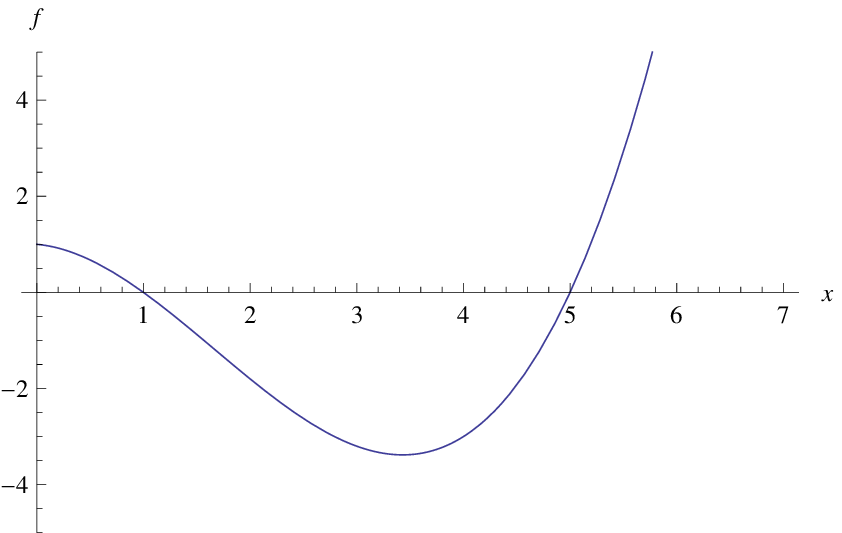}}                
  \subfloat[$\gamma=1$]{\includegraphics[width=0.3\textwidth]{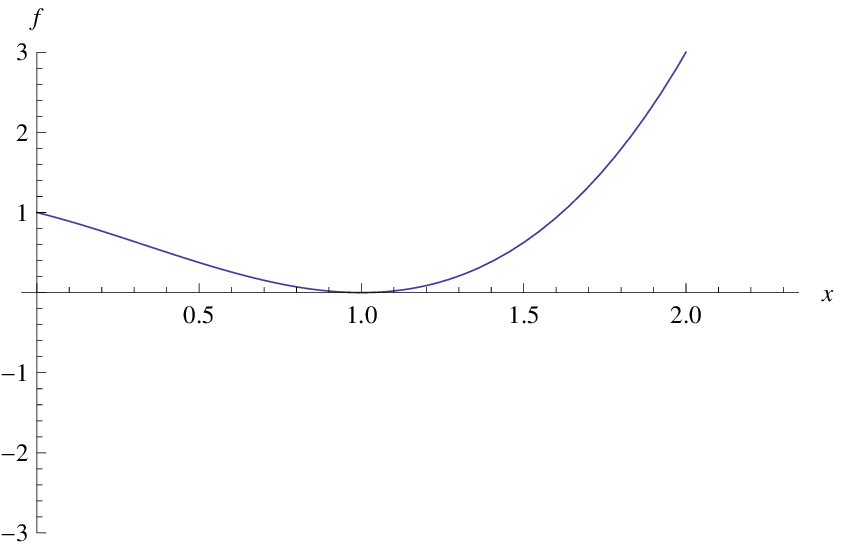}}
  \caption{$f(x)$ vs $x$}
\end{figure}

The solution to equation (\ref{new}) for $\gamma\leq 1$ is 
\begin{equation}
r^2=b^2\frac{1+\gamma(1-\rm{sn}^2(\textit{a},\textit{k}))}{\rm{sn}^2(\textit{a},\textit{k})} \qquad a=\sqrt{1+\gamma}\phi\qquad
k^2=\frac{2\gamma}{1+\gamma}.
\end{equation}
Note for this case, this trajectory is an exact expression as long as previous assumptions, namely a slowly varying metric, still hold.  In the limit $\beta\to\infty$, $\gamma\to 0$, $k\to 0$, $\rm{sn}\to \rm{sin}$,
\begin{equation}
r=\frac{b}{\rm{sin}(\phi)},
\end{equation}
just the equation of a strait line.  It is strait forward to calculate the first order angle of deflection\cite{landau}.
\begin{equation}
\Delta\phi=\frac{3\pi}{32}\gamma^{2}
\end{equation} 
\begin{figure}
  \centering
  \subfloat[$\gamma=0.5$]{\includegraphics[width=0.3\textwidth]{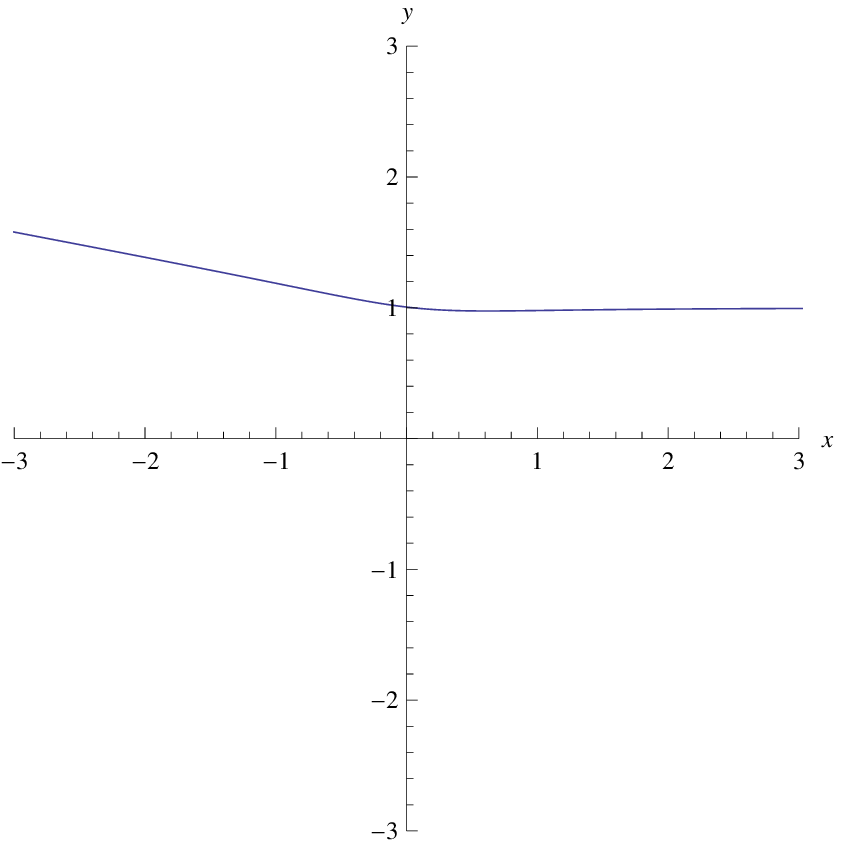}}                
  \subfloat[$\gamma=0.9$]{\includegraphics[width=0.3\textwidth]{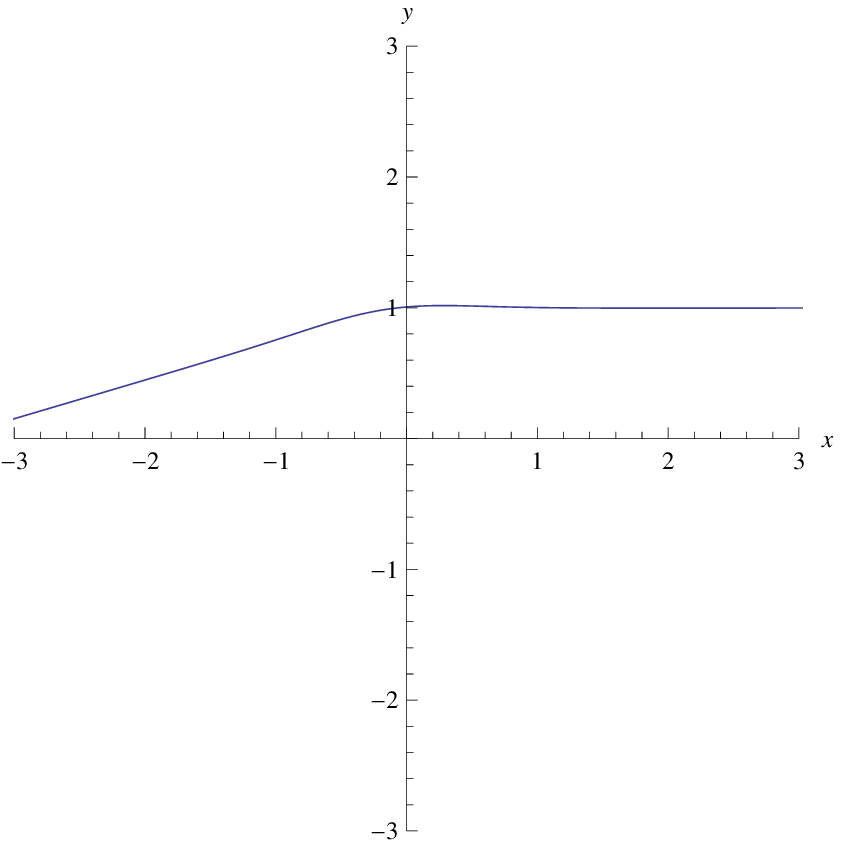}}
  \subfloat[$\gamma=1$]{\includegraphics[width=0.3\textwidth]{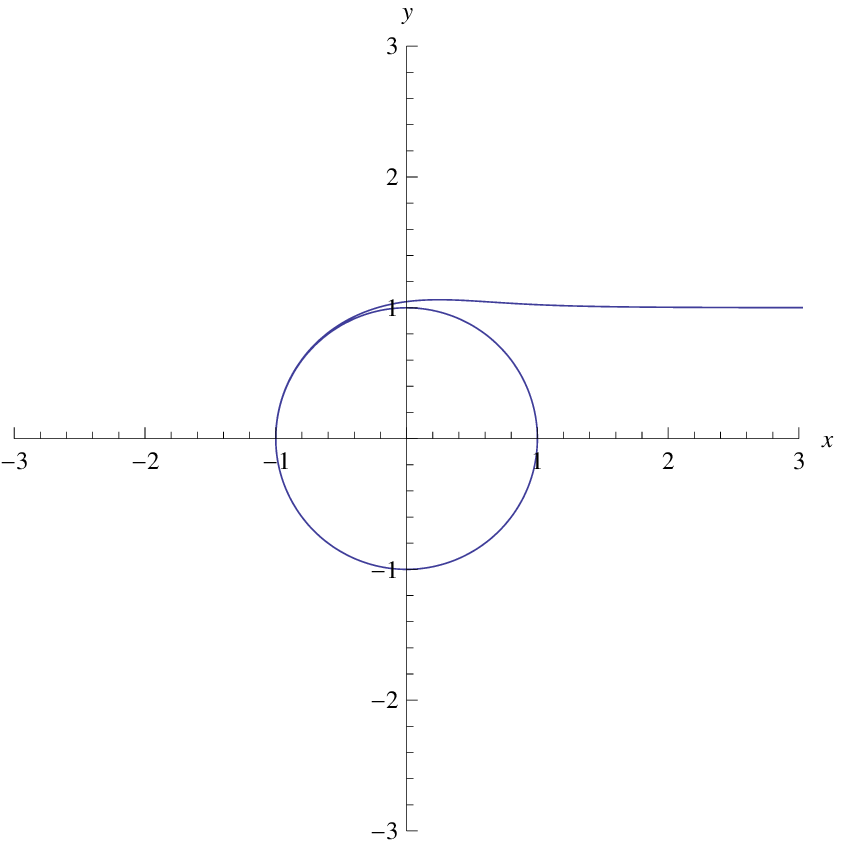}}
  \caption{Various Trajectories}
\end{figure}
For $\gamma$ approaching one from zero, the effective potential begins slightly repulsive, and then becomes more and more attractive until the ray is caught in circular orbit.  This orbit, however, is unstable. 

\section{Discussion}

This is a situation where closed time-like photon orbits can occur.  The Planck Field Strength is of the order $~10^{62}\,V/m$, but there is some discussion that possibly the string scale is much lower, possibly even to $~10^{46}\,V/m$\cite{ant}.  This is still an absurdly large value, much larger than the critical field strength of QED, $10^{18}\,V/m$.  Recall that this is the electric field strength such that there is an energy density of $m_e c^2$ per electron Compton wavelength cubed, $\lambda_e^3$.  At these energy densities, the electric field is primarily expected to decay into electron-positron pairs.  The behaviour of strong field QED needs to be better understood in order to know whether these field strengths can even be attained.    

This calculation shows behaviour of light that is usually only considered to arise from its interaction with gravity.  The Schwarchild solution reveals a similar unstable equilibrium in its potential function.  Only this time, it is not a gravitational force that is responsible for creating the closed orbit, but a non-linear electromagnetic mechanism.  This article presents another case where cosmological curiosities, such as black holes, event horizons, etc. that are usually confined to the study of gravity can be found in new places.

\section{Acknowledgements}

The author wishes to thank Dr. Gerardo Munoz and Dr. Emil Akhmedov for their helpful discussion and suggestions.

\end{document}